\documentclass[conference,letterpaper]{IEEEtran}

\usepackage{amsthm}
\usepackage[cmex10]{amsmath}
\usepackage{graphicx}
\usepackage[tight,footnotesize]{subfigure}
\usepackage{amsfonts}
\usepackage{algorithmic}
\usepackage{algorithm}
\usepackage{graphicx}
\usepackage{subfigure}
\usepackage{bm}
\usepackage{times,color,amssymb,epsfig,psfrag,stfloats,enumerate}
\usepackage{cite}
\usepackage{mathtools}
\usepackage{flushend}
\begin{document}
%
\title{Reducing Age-of-Information for Computation-Intensive Messages via Packet Replacement}

\author{\IEEEauthorblockN{Jie Gong\IEEEauthorrefmark{1}, Qiaobin Kuang\IEEEauthorrefmark{2}, Xiang Chen\IEEEauthorrefmark{2} and Xiao Ma\IEEEauthorrefmark{1}\\}
\IEEEauthorblockA{\IEEEauthorrefmark{1} School of Data and Computer Science, Sun Yat-sen University, Guangzhou 510006, China}
\IEEEauthorblockA{\IEEEauthorrefmark{2} School of Electronics and Information Technology, Sun Yat-sen University, Guangzhou 510006, China}
Email: gongj26@mail.sysu.edu.cn}


\maketitle

\begin{abstract}
Freshness of data is an important performance metric for real-time applications, which can be measured by \emph{age-of-information}. For computation-intensive messages, the embedded information is not available until being computed. In this paper, we study the age-of-information for computation-intensive messages, which are firstly transmitted to a mobile edge server, and then processed in the edge server to extract the embedded information. The packet generation follows \emph{zero-wait} policy, by which a new packet is generated when the last one is just delivered to the edge server. The queue in front of the edge server adopts \emph{one-packet-buffer replacement} policy, meaning that only the latest received packet is preserved. We derive the expression of average age-of-information for exponentially distributed transmission time and computing time. With packet replacement, the average age is reduced compared with the case without packet replacement, especially when the transmission rate is close to or greater than the computing rate.
\end{abstract}


%
\IEEEpeerreviewmaketitle

\section{Introduction}
Age-of-information, defined as the time elapsed since the generation of the latest delivered update, is one of the key metrics to measure the freshness of information in real-time monitoring and control applications \cite{kaul2012real}. Existing works mainly study the influence of queuing and transmission delay on the age performance. However, in computation-intensive applications such as autonomous driving and online facial recognition, an update, e.g. an image or a section of video record, needs to be processed to reveal the status information embedded in the packet. Due to the limited computational resource in the end devices, it is urgently required to adopt mobile edge computing (MEC) \cite{mao2017survey} technology to offload the computing tasks. This work focuses on the analysis of age-of-information for computation-intensive messages in MEC.

In the literature, age-of-information was initially studied in the elementary queuing systems such as $M/M/1, D/M/1$, and $M/D/1$ queuing models with first-come-first-served (FCFS) discipline \cite{kaul2012real}. As FCFS may result in large queuing delay, the last-come-first-served (LCFS) queue was considered to reduce the age-of-information \cite{kaul2012status}. Then, three packet management policies were introduced to further enhance the performance \cite{costa2016on}, where out-dated messages were discarded as they were less valuable for status update. Other than applying queuing analysis where update packets are generated randomly, update generation policies can be designed when the source has access to the channel's idle/busy state. The \emph{zero-wait} policy, which generates a fresh update just as the prior update is delivered and the channel becomes idle, was proposed in~\cite{yates2015lazy} to completely eliminate the waiting time in the queue. The optimality of the zero-wait policy was analyzed in~\cite{sun2017update}. The impact of computation on age has been recently considered in~\cite{alabbasi2018joint} which focuses on scheduling in computation and networking with centralized cloud. Nevertheless, it is still an open problem to characterize age-of-information with computing in MEC.

In MEC, each packet experiences two stages: transmission and computing, which can be viewed as a two-hop network. Among the multi-hop related research efforts, the optimality of the last-generated-first-served (LGFS) policy was analyzed in multi-hop networks~\cite{bedewy2017age}. The age-of-information for multi-flow multi-hop networks with interference was studied in~\cite{talak2017minimizing}. In the multi-hop line network with preemptive servers and random arrivals, a simple expression of average age was obtained in \cite{yates2018age, yates2018the} using stochastic hybrid systems (SHS) tool. Different from existing works, our preliminary work~\cite{kuang2019age} derived the expression for the system with zero-wait policy in transmission stage and $M/M/1$ FCFS queue in computing stage. In this paper, we further consider packet replacement policy in computing stage. In particular, the edge server preserves a queue of length one. If the queue is full and a new packet arrives at the server, the old packet in the queue is discarded and replaced by the new one. We characterize the age-of-information by deriving the distribution of transmission time, waiting time and inter-arrival time of the successfully computed packets. Numerical results illustrate that the message going through the system with packet replacement is fresher than that without packet replacement.

\section{System Model}
\begin{figure}[t]
\centering
\includegraphics[width=3.0in]{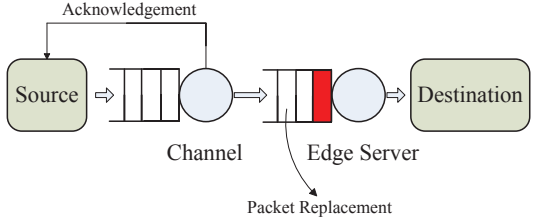}
\caption{Status update system with MEC.} \label{fig:system}
\end{figure}

Consider a status update system for computation-intensive messages which are processed at mobile edge server as shown in Fig.~\ref{fig:system}. The whole procedure is divided into two stages: \emph{transmission stage} and \emph{computing stage}. In the transmission stage, an update packet is firstly generated from the source, and then transmitted through the channel. The generation of update packets follows zero-wait policy, by which a new update is generated by the source when the transmission of the previous update is just completed. Therefore, there is no waiting queue in the transmission stage. In the computing stage, the packet is received and computed by the edge server so that the information embedded in the packet is revealed to the destination. Both transmission and computing times are assumed to be random and follow exponential distribution with means $1/\lambda$ and $1/\mu$, respectively, leading to queuing in the computing stage. In this paper, we consider a \emph{one-packet-buffer replacement queue}, i.e., at most one packet is allowed to wait in the queue, and it is replaced if a new one arrives. In this way, the out-dated update is discarded and the freshest update goes into the queue for computing. It is expected to reduce the age-of-information compared with FCFS queue.

Denote $\tau_i$ as the generation time instant of the $i$-th update packet, $i = 1, 2, \cdots$.  As zero-wait policy is adopted, $\tau_i$ is also the transmission completion time of the $(i\!-\!1)$-th packet. Notice that not all the packets are received by the destination due to the packet replacement in the computing stage. We mainly focus on the computed packets as they account for the average age-of-information. For ease of analysis, the computed packets are re-indexed by $k = 1, 2, \cdots$. Denote $t_k$ as the completion time instant of the transmission stage for the $k$-th computed packet, and $t_k'$ as the completion time instant of the computing stage for the $k$-th computed packet.

At the time instant $t$, the index of the latest information received by the destination is denoted by
\begin{align}
K(t) \coloneqq \max \{ k|t_k' \le t\}. \label{eq:kt}
\end{align}
As $\tau_{i(t)} = t_{K(t)}$ for some $i(t)$, the generation time instant of this packet can be given as
$
U(t) \coloneqq \tau_{i(t)-1}.
$
The age-of-information is defined as
\begin{align}
\Delta(t) \coloneqq t - U(t).
\end{align}

\begin{figure}[t]
\centering
\includegraphics[width=3.4in]{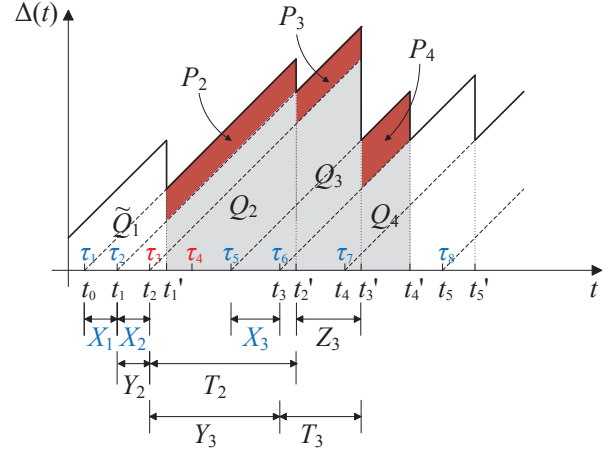}
\caption{A sample path of age-of-information with MEC.} \label{fig:aoi}
\end{figure}

The sawtooth shaped sample path of the random process $\Delta(t)$ is illustrated in Fig.~\ref{fig:aoi}. It can be seen from the figure that since the fifth packet has already arrived at the edge server just before the second packet is successfully computed. Hence, the fifth packet is re-indexed as the third computed packet.

The average age in the observation range $(0, t)$ is defined by
\begin{align}
\Delta_{t} \coloneqq \frac{1}{t}\int_0^{t} \Delta(\tau)\mathrm{d} \tau.
\end{align}
The integration is equal to the area below the curve $\Delta(t)$, which can be calculated as the summation of the areas of parallelograms $P_k$ and trapezoids $Q_k, k = 2, \cdots, K(t)$, where $K(t)$ is defined in \eqref{eq:kt}. Therefore, the average age can be written as
\begin{align}
\Delta_{t} &= \frac{K(t)-1}{t} \times \frac{\tilde Q_1 + \sum_{k=2}^{K(t)} (P_k + Q_k) + Q^*}{K(t)-1}, \label{eq:avg_aoi}
\end{align}
where $Q^*$ is the area in time interval $(t_{K(t)}', t)$.
As $t$ goes to infinity, $K(t)$ goes to infinity as well. Consequently, the term $\frac{\tilde Q_1 + Q^*}{K(t)-1}$ goes to zero as the nominator is finite. To calculate $P_k$ and $Q_k$, denote
$
X_k \coloneqq \tau_i-\tau_{i-1}
$
as the transmission time for the $k$-th computed packet, where $i$ satisfies $\tau_i = t_k$,
$
Y_k \coloneqq t_k - t_{k-1}
$
as the time elapsed between the transmission completion time instants of the $k$-th and $(k\!-\!1)$-th computed packets,
$
T_k \coloneqq t_k' - t_k
$
as the system time of the $k$-th computed packet in the computing stage, including waiting time and service time, and
$
Z_k \coloneqq t_k' - t_{k-1}'
$
as the inter-departure time of the computing stage. We have
\begin{align}
P_k &= X_{k-1}Z_k, \label{eq:pk}\\
Q_k &= \frac{1}{2}(T_k + Y_k)^2 - \frac{1}{2}T_{k-1}^2, \label{eq:qk}\\
t &= t_1  + \sum_{k=2}^{K(t)} Y_k + t^*,\label{eq:tao}
\end{align}
where $t^* = t - t_{K(t)}'$. Substituting \eqref{eq:pk}-\eqref{eq:tao} into \eqref{eq:avg_aoi}, and letting $t$ going to infinity, we have
\begin{align}
\bar{\Delta} &= \lim_{t \rightarrow +\infty} \Delta_{t} = \frac{E[P_k] + E[Q_k]}{E[Y_k]} \nonumber\\
&= \frac{1}{E[Y_k]} \left( E[X_{k-1}Z_k] + E[T_kY_k] + \frac{1}{2}E[Y_k^2] \right), \label{eq:bar_aoi}
\end{align}
where $E[\cdot]$ is the expectation operator, and the fact that $E[T_k^2] = E[T_{k-1}^2]$ is used.

\emph{Remark 1:} The inter-departure time $Z_k$ is independent of $X_{k-1}$ as it only depends on the arrival process at the edge server between the $(k\!-\!1)$-th and $k$-th packets and the system times of these two packets in computing stage. The influence of $X_{k-1}$ to $Z_k$ is blocked by the one-packet-buffer replacement queue principle. Hence, we have \begin{align}
E[X_{k-1}Z_k] = E[X_{k-1}]E[Z_k].
\end{align}
Furthermore, we have by definition that
\begin{align}
E[Z_k] = \!\lim_{t \rightarrow +\infty}\! \frac{\sum_{k=2}^{K(t)} \!Z_k}{K(t)\!-\!1} = \!\lim_{t \rightarrow +\infty}\! \frac{t}{K(t)\!-\!1} = E[Y_k]. \label{eq:Zk}
\end{align}
Hence, to calculate the average age-of-information, we only need to consider $X_k$, $Y_k$ and $T_k$.

\emph{Remark 2:} The average area $E[Q_k]$ is equal to the one in \cite{costa2016on} for $M/M/1/2^*$ queue that is derived based on whether the queue is empty or not upon departure. For completeness of description, we re-calculate the result by proposing another method, which directly derives the distributions of $X_k$ and $Y_k$, which are detailed in the next section.

\section{Calculation of Average Age}
In this section, we derive the average age according to \eqref{eq:bar_aoi}. By definition, we have
$
T_k = W_k + S_k,
$
where $W_k$ and $S_k$ are the waiting time and the service time in the computing stage, respectively. As $X_k$ and $Y_k$ depend on the system time of the $(k\!-\!1)$-th packet, we firstly derive the distribution of $W_k$. Then, we calculate the terms in \eqref{eq:bar_aoi} one by one.

\subsection{Distribution of $W_k$}
The event that the $k$-th packet has zero waiting time occurs if and only if there is no arrival during the service time $s$ of the $(k\!-\!1)$-th packet, which is equivalent to the event that the inter-arrival time in computing stage is larger than $s$. Since the inter-arrival time of the computing stage is exponentially distributed, we have
\begin{align}
\mathrm{Pr}(W_k = 0) &= \int_0^{\infty} \mathrm{Pr}(\textrm{no~arrival~in~}(0,s)) f_{S_{k-1}}(s) \mathrm d s \nonumber\\
&= \int_0^{\infty} e^{-\lambda s} \mu e^{-\mu s} \mathrm d s = \frac{\mu}{\lambda + \mu}. \label{eq:W0}
\end{align}

For the case that $W_k > 0$, We consider the probability $\mathrm{Pr}(0 < W_k \le w | S_{k-1} = s)$. Notice that in the computing stage, the waiting time of the $k$-th packet is no longer than the service time of the $(k\!-\!1)$-th packet $s$. If $s \le w$, $W_k \le w$ is guaranteed. Then, the event $W_k > 0$ happens if and only if at least one packet arrives in time duration of length $s$. If $s > w$ on the other hand, the event $0 < W_k \le w$ happens if and only if at least one packet arrives in time duration $(s, w+s)$\footnote{In this notation, we reset the transmission completion time instant of the $(k\!-\!1)$-th  computed packet as 0. This is valid throughout this section.}. According to the total probability formula, we have
\begin{align}
&{~}\mathrm{Pr}(0 < W_k \le w) \nonumber\\
=&{~} \int_0^{\infty} \mathrm{Pr}(0 < W_k \le w | S_{k-1} = s) f_{S_{k-1}}(s) \mathrm d s \nonumber\\
=&{~} \int_0^w (1-e^{-\lambda s})\mu e^{-\mu s} \mathrm d s + \int_w^{\infty} (1-e^{-\lambda w}) \mu e^{-\mu s} \mathrm d s \nonumber\\
=&{~} \frac{\lambda}{\lambda+\mu} (1-e^{-(\lambda+ \mu)w}).
\end{align}
Hence, for $w > 0$, the probability density function of $W_k$ is
\begin{align}
f_{W}(w) \coloneqq f_{W_k}(w) = \lambda e^{-(\lambda+ \mu)w}, \quad w > 0. \label{eq:W1}
\end{align}

\subsection{Distribution of $X_k$}
Recall that $X_k$ is the transmission time for the $k$-th computed packet. As $X_k$ is related to the waiting and computing process of the $(k\!-\!1)$-th packet, we derive the distribution of $X_k$ conditioned on $W_{k-1}$ and $S_{k-1}$ by analyzing the probability $\mathrm{Pr}(X_k > x|W_{k-1} = w, S_{k-1} = s)$. Given $W_{k-1}$ and $S_{k-1}$, we analyze the conditions on which the event $X_k > x$ occurs. In general, if the $k$-th packet arrives at time instant $t$, $X_k > x$ occurs if there are no packet arrivals before $t$ for a time duration longer than $x$ and no arrivals after $t$ until the computing completion time for the $(k\!-\!1)$-th packet. The detailed results are as follows:

\subsubsection{$0<w \le s$} If $0<w \le s$, and $x \le w$, the event $X_k > x$ occurs when the $k$-th packet arrives in the small interval $(t, t+\mathrm d t)$ while at the same time no packet arrivals during time intervals $(t-x, t)$ and $(t+\mathrm d t, w+s)$, or no packet arrivals during $(w, w+s)$. The probability that a single packet arrives in $(t, t+\mathrm d t)$ is $\lambda \mathrm d t + o(\mathrm d t)$, Hence, the probability $\mathrm{Pr}(X_k > x|W_{k-1} = w, S_{k-1} = s)$ is the integral over all possible $t$, i.e.,
\begin{align}
&{~}\mathrm{Pr}(X_k > x|W_{k-1} = w, S_{k-1} = s) \nonumber\\
=&{~} \int_{w}^{w+x} e^{-\lambda (t-w)} e^{-\lambda(w+s-t)} \lambda \mathrm d t \nonumber\\
&\quad + \int_{w+x}^{w+s} e^{-\lambda x} e^{-\lambda(w+s-t)} \lambda \mathrm d t + e^{-\lambda s} \label{eq:Prx1}\\
=&{~} x \lambda e^{-\lambda s} + e^{-\lambda x}. \label{eq:F1}
\end{align}
Notice that we ignore $e^{\lambda \mathrm d t}$ and $o(\mathrm d t)$ as they are higher-order infinitesimal. The first integral in \eqref{eq:Prx1} refers to the special case that the time interval before $t$ is shorter than $x$. As there are no packet arrivals during $(0, w)$ by definition, this special case also results in $X_k > x$.

If $w < x \le s$, the analysis is similar to the above, but the integral range is from $x$ to $w+s$. Hence, we have
\begin{align}
&{~}\mathrm{Pr}(X_k > x|W_{k-1} = w, S_{k-1} = s) \nonumber\\
=&{~} \int_{x}^{w+x} e^{-\lambda (t-w)} e^{-\lambda(w+s-t)} \lambda \mathrm d t \nonumber\\
&\quad + \int_{w+x}^{w+s} e^{-\lambda x} e^{-\lambda(w+s-t)} \lambda \mathrm d t + e^{-\lambda s} \\
=&{~} w \lambda e^{-\lambda s} + e^{-\lambda x}. \label{eq:F2}
\end{align}

If $s < x \le w+s$, the event $X_k > x$ occurs when the $k$-th packet arrives in $(t, t+\mathrm d t)$ while at the same time no packet arrivals during time intervals $(w, t)$ and $(t+\mathrm d t, w+s)$ for $x<t<w+s$, or no packet arrivals during $(w, w+s)$. We have
\begin{align}
&{~}\mathrm{Pr}(X_k > x|W_{k-1} = w, S_{k-1} = s) \nonumber\\
=&{~} \int_{x}^{w+s} e^{-\lambda (t-w)} e^{-\lambda(w+s-t)} \lambda \mathrm d t + e^{-\lambda s} \nonumber\\
=&{~} (w + s -x) \lambda e^{-\lambda s} + e^{-\lambda s}. \label{eq:F3}
\end{align}

Finally, for the case that $x > w+s$, the event $X_k > x$ occurs when there are no arrival during time interval $(w, x)$. Therefore,
\begin{align}
\mathrm{Pr}(X_k > x|W_{k-1} = w, S_{k-1} = s)  = e^{-\lambda(x-w)}. \label{eq:F4}
\end{align}

\subsubsection{$s < w$} In this case, we analyze the probability in the same way. If $x \le s$, it is easy to verify that the result is equal to \eqref{eq:F1}. Similarly, the result with $w < x \le w+s$ is the same as \eqref{eq:F3}, and the result with $x > w+s$ is the same as \eqref{eq:F4}. While for $s < x \le w$, the event $X_k > x$ occurs when the $k$-th packet arrives in $(t, t+\mathrm d t)$ while at the same time no packet arrivals during time intervals $(w, t)$ and $(t+\mathrm d t, w+s)$ for $w < t < w+s$, or no packet arrivals during $(w, w+s)$. We have
\begin{align}
&{~}\mathrm{Pr}(X_k > x|W_{k-1} = w, S_{k-1} = s) \nonumber\\
=&{~} \int_{w}^{w+s} e^{-\lambda (t-w)} e^{-\lambda(w+s-t)} \lambda \mathrm d t + e^{-\lambda s} \nonumber\\
=&{~} s \lambda e^{-\lambda s} + e^{-\lambda s}.
\end{align}

\subsubsection{$w = 0$} This case can be viewed the extreme case of $w \le s$. By setting $w=0$ in \eqref{eq:F1}-\eqref{eq:F4} and check the validity, we have
\begin{align}
\mathrm{Pr}(X_k > x|W_{k-1} = 0, S_{k-1} = s) = e^{-\lambda x}.
\end{align}

In summary, by taking the derivative of $\mathrm{Pr}(X_k > x|W_{k-1} = w, S_{k-1} = s)$ in terms of $x$, we can obtain the conditional probability density function
\begin{align}
&{~}f_{X|W, S}(x|w, s) \coloneqq f_{X_k|W_{k-1}, S_{k-1}}(x|w, s) \nonumber\\
=&{~} \left\{\begin{array}{ll} \lambda e^{-\lambda x} - \lambda e^{-\lambda s}, & x \le \min\{w, s\}, \\
 \lambda e^{-\lambda x}, & w < x \le s, \\
 \lambda e^{-\lambda s}, & \max \{ w, s\} < x \le w+s, \\
 \lambda e^{-\lambda (x-w)}, & x > w+s, \\
 0, & \textrm{else}.
 \end{array} \right. \label{eq:xk}
\end{align}

According to \eqref{eq:W0}, \eqref{eq:W1} and \eqref{eq:xk}, by the law of total expectation, we have
\begin{align}
&\quad\; E[X_k] \nonumber\\
&= E[E[X_k|W_{k-1}, S_{k-1}]] \nonumber\\
&= \mathrm{Pr}(W_k = 0) \int_0^{\infty} \mu e^{-\mu s} E[X_k|W_{k-1} = 0, S_{k-1} = s] \mathrm d s \nonumber\\
&\; + \!\int_0^{\infty} \!\lambda e^{-(\lambda+ \mu)w} \!\int_0^{\infty} \!\mu e^{-\mu s} E[X_k|W_{k\!-\!1} \!=\! w, S_{k\!-\!1} \!=\! s] \mathrm d s \mathrm d w \nonumber\\
&= \mathrm{Pr}(W_k = 0) \int_0^{\infty} \mu e^{-\mu s} \int_0^{\infty} x \lambda e^{-\lambda x} \mathrm d x \mathrm d s \nonumber\\
&\; + \!\int_0^{\infty} \!\lambda e^{-(\lambda+ \mu)w} \!\int_0^{\infty} \!\mu e^{-\mu s} \!\int_0^{\infty} x f_{X|W, S}(x|w, s) \mathrm d x \mathrm d s \mathrm d w \nonumber\\
&= \frac{1}{\mu} \left(1+ \frac{1}{\rho(1+\rho)} - \frac{\rho^3}{(1+\rho)^4} - \frac{\rho^2}{(1+\rho)^2} \right), \label{eq:EXk}
\end{align}
where $\rho = \lambda/\mu$, and the second integral is calculated by dividing the integral region based on \eqref{eq:xk}.

\subsection{Distribution of $Y_k$}
Recall that $Y_k$ is the time elapsed between the transmission completion time instants of the $k$-th and $(k\!-\!1)$-th computed packets. Similar to the previous subsection, we derive the conditional probability density function $f_{Y_k|W_{k-1}, S_{k-1}}(y|w,s)$ by calculating the conditional probability $\mathrm{Pr}(Y_k \le y|W_{k-1} = w, S_{k-1} = s)$. When $w \le y < w+s$, the event $Y_k \le y$ occurs if and only if there is at least one packet arrival in $(w, y)$ and no packet arrivals in $(y, w+s)$, i.e.,
\begin{align}
&{~}\mathrm{Pr}(Y_k \le y|W_{k-1} = w, S_{k-1} = s) \nonumber\\
=&{~} (1-e^{-\lambda(y-w)})e^{-\lambda(w+s-y)} \nonumber\\
=&{~} e^{-\lambda(w+s-y)} - e^{-\lambda s}.
\end{align}

When $y \ge w+s$, the event $Y_k > y$ happens if and only if no packet arrivals in $(w, y)$. We have
\begin{align}
&{~}\mathrm{Pr}(Y_k \le y|W_{k-1} = w, S_{k-1} = s) \nonumber\\
=&{~} 1- \mathrm{Pr}(Y_k > y|W_{k-1} = w, S_{k-1} = s)\nonumber\\
=&{~} 1 - e^{-\lambda(y-w)}.
\end{align}

By taking derivation of $\mathrm{Pr}(Y_k \le y|W_{k-1} = w, S_{k-1} = s)$, we have
\begin{align}
f_{Y|W, S}(y|w,s) &\coloneqq f_{Y_k|W_{k-1}, S_{k-1}}(y|w,s) \nonumber\\
&= \left\{ \begin{array}{ll} \lambda e^{-\lambda(w+s-y)}, & w \le y < w+s, \\
 \lambda e^{-\lambda(y-w)}, & y \ge w+s. \end{array} \right.
\end{align}

To calculate $E[Y_k]$ and $E[Y_k^2]$, we derive the probability density function of $Y_k$ as
\begin{align}
f_{Y_k}(y) &= \mathrm{Pr}(W_{k-1} = 0) \int_0^{\infty} f_{Y|W, S}(y|0,s) \mu e^{-\mu s} \mathrm d s \nonumber\\
 & \;  +\int_0^{\infty} \lambda e^{-(\lambda+ \mu)w}  \int_0^{\infty} f_{Y|W, S}(y|w,s) \mu e^{-\mu s} \mathrm d s \mathrm d w \nonumber\\
&= \left( \frac{\lambda^2}{\mu} + \frac{\lambda\mu}{\lambda+\mu}\right) e^{-\lambda y} + \frac{\lambda \mu (\lambda + 2 \mu)}{(\lambda + \mu)^2} e^{- \mu y} \nonumber\\
&\;- \left( \frac{\lambda^2}{\mu} + \frac{2\lambda\mu}{\lambda+\mu} + \lambda^2 y\right) e^{-(\lambda + \mu) y}.
\end{align}

Accordingly, we can obtain
\begin{align}
E[Y_k] &= \frac{1}{\mu}\frac{1 + \rho + \rho^2}{\rho (1 + \rho)}, \label{eq:EYk}\\
E[Y_k^2] &= \frac{2}{\mu^2} \left(1+ \frac{1}{\rho^2} - \frac{\rho(1 + 2\rho)}{(1 + \rho)^4}\right). \label{eq:EYksquare}
\end{align}

\subsection{Calculation of $E[T_kY_k]$}

To calculate $E[T_kY_k]$, we rewrite $T_k$ as a function of $W_{k-1}$ and $S_{k-1}$. We observe the relation between $T_{k-1}$ and $Y_k$. If $T_{k-1} > Y_k > W_{k-1}$, i.e., the $k$-th computed packet arrives at the edge server when the $(k\!-\!1)$-th packet is still being processed, we have $W_k = T_{k-1} - Y_k$. Otherwise, $W_k = 0$. Hence, we have
\begin{align}
W_k = (T_{k-1} - Y_k)^+ = (W_{k-1} + S_{k-1} - Y_k)^+.
\end{align}

Therefore, we obtain
\begin{align}
E[T_kY_k] &= E[(W_k + S_k)Y_k] \nonumber\\
&= E[(W_{k-1} + S_{k-1} - Y_k)^+Y_k] + E[S_k]E[Y_k] \label{eq:TkYk}
\end{align}
as $S_k$ and $Y_k$ are independent with each other. By utilizing the distributions of $W_{k-1}$ and $S_{k-1}$, and the conditional probability density function $f_{Y|W, S}(y|w, s)$, we can calculate that
\begin{align}
&\quad\;E[(W_{k-1} + S_{k-1} - Y_k)^+Y_k] \nonumber\\
&= \mathrm{Pr}(W_{k-1} = 0) \int_0^{\infty} \mu e^{-\mu s} \int_0^{s} (s-y)y \lambda e^{-\lambda(s-y)} \mathrm d s \mathrm d y \nonumber\\
 & \;  +\int_0^{\infty} \lambda e^{-(\lambda+ \mu)w}  \int_0^{\infty} \mu e^{-\mu s} \nonumber\\
 &\qquad\qquad \cdot \int_w^{w+s} (w+s-y)y \lambda e^{-\lambda(w+s-y)} \mathrm d y \mathrm d s \mathrm d w \nonumber\\
&= \frac{1}{\mu^2}\left( \frac{1}{1+\rho} - \frac{1+2\rho}{(1+\rho)^4}\right). \label{eq:WkYk}
\end{align}

Since $E[S_k] = 1/\mu$, summarizing \eqref{eq:EYk}, \eqref{eq:TkYk} and \eqref{eq:WkYk}, we have
\begin{align}
E[T_kY_k] = \frac{1}{\mu^2}\left( 1 + \frac{1}{\rho} - \frac{1+2\rho}{(1+\rho)^4}\right). \label{eq:ETkYk}
\end{align}

\subsection{Average Age-of-Information}
Finally, according to \eqref{eq:bar_aoi}-\eqref{eq:Zk}, \eqref{eq:EXk}, \eqref{eq:EYk}, \eqref{eq:EYksquare} and \eqref{eq:ETkYk}, we obtain
\begin{align}
\bar{\Delta} = \frac{1}{\mu} \left( 2 + \frac{2}{\rho} + \frac{1+3\rho}{(1+\rho)^2} - \frac{\rho^3}{(1+\rho)^4} - \frac{2(1+\rho)}{1+\rho + \rho^2}\right).
\end{align}

\section{Numerical Comparison}

\begin{figure}[t]
\centering
\includegraphics[width=3.4in]{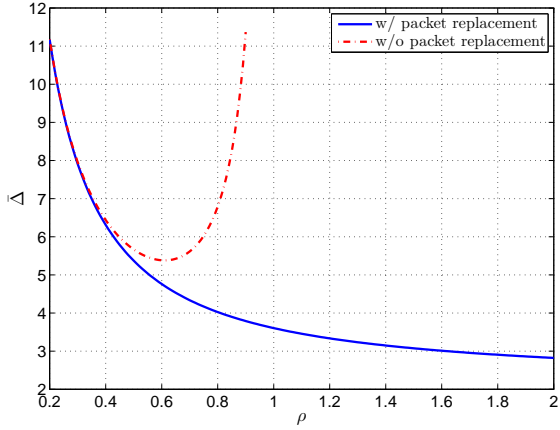}
\caption{Performance comparison for status update via mobile edge computing with or without packet replacement. $\mu = 1$.} \label{fig:compare}
\end{figure}

In this section, the average age-of-information for computation-intensive messages with packet replacement is compared with that without packet replacement as in \cite{kuang2019age}, where all the packets wait in the queue for processing with FCFS discipline. As shown in Fig.~\ref{fig:compare}, with packet replacement, the average age-of-information decreases as $\rho$ increases. The phenomenon is reasonable as with the increase of the channel transmission rate, the update packet waiting for computing is fresher as it is replaced by the latest update faster. Asymptotically, the minimum average age is achieved when $\rho \rightarrow +\infty$, which results in $\bar{\Delta}_{\min} = 2/\mu$. In comparison, the average age without replacement first decreases as $\rho$ increases, and then increases to infinity as $\rho \rightarrow 1$. When $\rho$ is close to 1, the queue length in edge server becomes quite long, and the age becomes large due to the long time waiting in the queue.

\section{Conclusion and Future Work}
In this paper, we have derived the average age-of-information for two-stage mobile computing system with zero-wait and packet replacement. The stationary distributions of some random processes are obtained, including the waiting time $W_k$ before being computed, the transmission time $X_k$ for the computed packet, and the inter-arrival time $Y_k$ of two consecutive computed packets. It is shown that with packet replacement, the average age is reduced compared with the case without packet replacement, and the value tends to a minimum $2/\mu$ when the transmission rate tends to infinity. Future work includes finding other packet generation policies instead of zero-wait to further reduce the average age, and consider the cases with multiple users or multiple edge servers.

\bibliographystyle{IEEEtran}
\bibliography{ref}

\end{document}